\documentclass[aps,prl,twocolumn,showpacs]{revtex4-1}  
\usepackage{graphicx} 
\usepackage{amssymb}
\usepackage{float}
\usepackage{amsfonts,amssymb,amsmath,amsgen,amsopn,amsbsy,theorem,graphicx,epsfig}
\usepackage[utf8]{inputenc}
\usepackage[english]{babel}
\begin{document}
\title{Fano enhancement of second harmonic field via dark-bright plasmon coupling}
\author{ Mehmet G\"{u}nay}\thanks{gunaymehmt@gmail.com}
\affiliation{Department of Nanoscience and Nanotechnology, Faculty of Arts and Science, \\
 Burdur Mehmet Akif Ersoy University, 15030 Burdur, Turkey}
\begin{abstract}
Surface plasmon resonances, the coherent oscillation of free electrons, can concentrate incident field into small volumes much smaller than the incident wavelength. The intense fields at these \textit{hot spots} enhance the light-matter interactions and may lead to the appearance of nonlinearity. Controlling such nonlinearities is significant for various practical applications. Here we report that by coupling dark modes to the first and the generated second harmonic modes separately, one can gain control over both fields. We find that by engineering path interferences (Fano resonances) between bright and dark plasmon modes it is possible to enhance the fundamental mode without increasing the nonlinear field, enhance the nonlinear field without modifying the fundamental mode, and enhance the second harmonic field with enhanced fundamental mode.
\end{abstract}
\maketitle

\section{Introduction} \label{Intro}
Controlling the interaction between light and matter is fundamental to science, as well as to the technological perspective; ranging from probing entanglement in quantum physics to controlled use of the spectacular information-carrying capacity. Traditionally light can only be controlled on length scales down to a little below the wavelength of light. On the other hand, plasmonic nanoparticles can be used to confine light to subwavelength dimensions below the diffraction limit to achieve photonic circuit miniaturization~\citep{Bozhevolnyi}. Fano resonances, an analog of electromagnetically induced transparency, in these structures are of importance. Such type of devices can be probe experimentally and can be useful in slow-light propagation~\citep{Huang,Wu}, electromagnetic induced absorption~\citep{Tassin,Taubert}, an anomaly in light transmission~\citep{Imura}, and solar cell~\citep{Kauranen} applications.

Recent years have witnessed significant progress in the field of quantum plasmonics. Unlike plasmonics, quantum plasmonics involves the study of the quantum properties of light and its interaction with the matter at the nanoscale. It provides a unique platform for the manipulation of light through its interaction with electromagnetic fields, well below the diffraction limit~\citep{1,2}. Placing an optical emitter near a metal nanostructure can lead to observe light-matter interactions at different levels, i.e., from weak to ultra-strong coupling limit~\citep{16}. The emitters used can be a quantum dot~\citep{11}, a molecule~\citep{12}, and a nitrogen-vacancy~(NV) center~\citep{13}. Fano resonances, observed in the weak coupling regime~\citep{15}, can increase the lifetime of plasmon excitation~\citep{24}, which leads to further enhancement of the hot spot field. This extra enhancement~\citep{21} enables the operation of the plasmonic laser (SPASER)~\citep{3} and gives rise to the enhancement of nonlinear processes, such as second-harmonic generation~(SHG)~\citep{22} and four-wave mixing~(FWM)~\citep{23}. Rabi splitting in the strong coupling regime~\citep{16,17}, however, modifies the resonance scheme and creates two distinct modes, which can be used as an optical switch~\citep{20}.

\begin{figure*}[h]
\includegraphics[width=12cm]{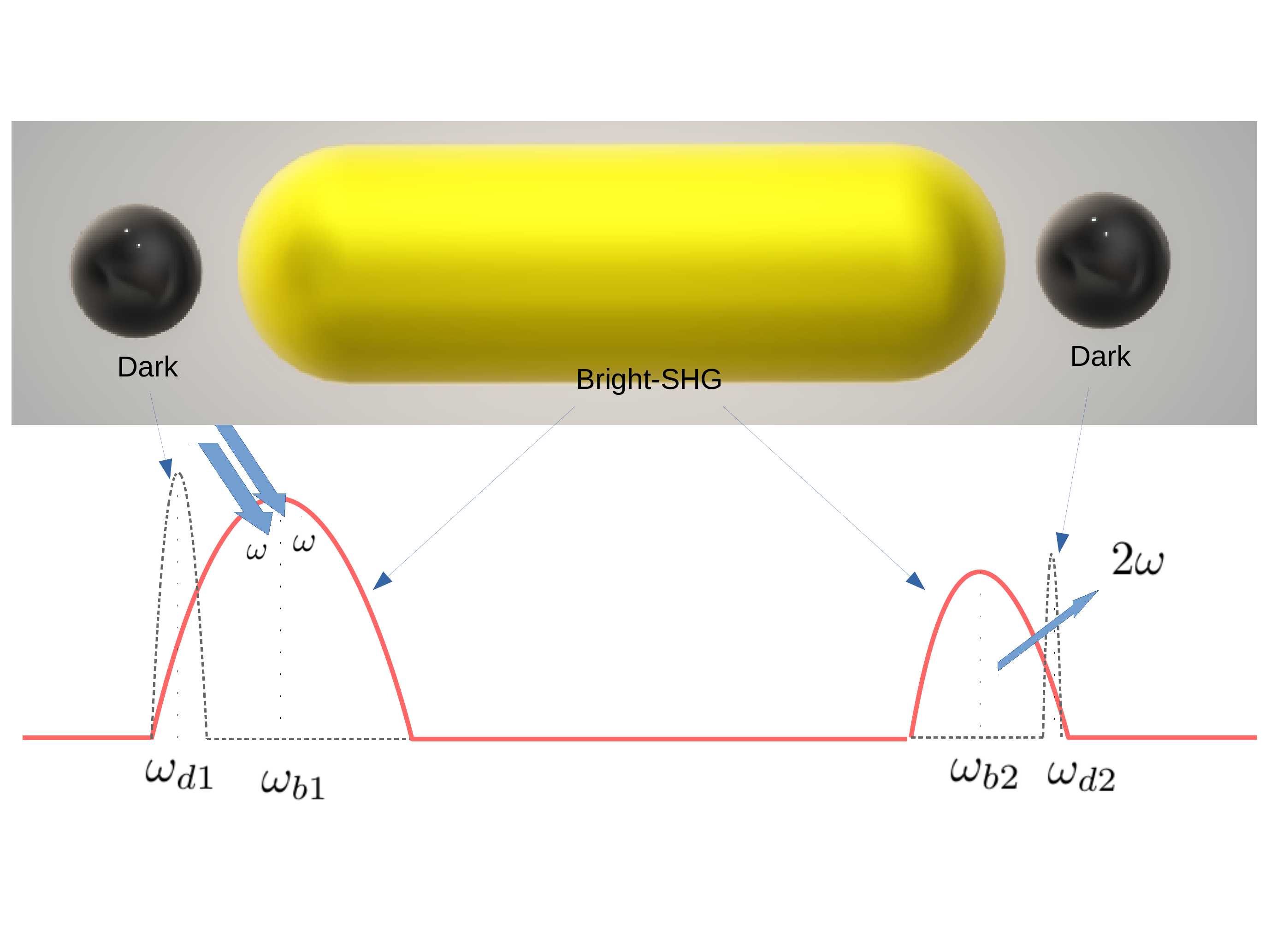}
\caption{Illustration of the metal nano-spheres/nanorod hybrid system considered. Two nano-spheres (black) support plasmonic dark modes coupled to the metal nanorod (top). Two $\omega$ plasmons combine to generate 2$\omega$ plasmons. Left (right) dark spheres having resonance at $\omega_{d1}$~($\omega_{d2}$) interacts with the FH (SH) mode of the nanorod~(bottom). }
\label{fig1}
\end{figure*}

Enhancement, due to path interference effects, of both linear~\citep{18} and nonlinear~\citep{19} phenomena that emerged from the hot spots of plasmonic structures, is well studied via coupling plasmonic modes to the two-level quantum structures. These path interference effects (Fano resonances), besides the emitter-plasmon coupling, can also appear when the plasmon mode of the metal nanoparticle is coupled to a long-live dark plasmon mode~\citep{14}. The latter can be advantageous in device applications since the performance of the former ones can decay in time. In this paper, we utilize bright-dark plasmon coupling in which to gain control over both the linear and the nonlinear field intensity. Different from our previous work~\citep{19}, here, we consider both linear and nonlinear Fano resonances at the same setup by coupling two distinct plasmonic dark modes to the metal nanorod having second harmonic generation property, as shown in Fig.~\ref{fig1}.

It is known that enhancement of the second harmonic field can be observed by enhancing theIn this paper,  first harmonic mode resulting in the enhanced second-harmonic field (linear Fano resonance) or coupling the emitter to the second harmonic mode (nonlinear Fano resonance) to enhance the second harmonic field without modifying the first harmonic field~\citep{19}. The existence of the second long-live mode, as studied here, can be useful to enhance one of the modes by suppressing the other or to obtain the better enhancement of the nonlinear field intensity by applying both enhancement schemes at the same time. 

In the following, we first present a basic analytical model in Sec.~\ref{model}, which is demonstrated to treat Fano resonances realistically and explain how to control the intensity of each field. It becomes apparent that the path interferences in the linear and nonlinear Fano resonances can be controlled separately, where one can enhance or suppress the nonlinear field intensity while controlling the first harmonic mode by carefully chosen nanoparticles. In Sec.~\ref{Conclusion}, we discuss the possible implementations and summarize the findings.

\section{The Model} \label{model}

We consider the structure as shown in Fig.~\ref{fig1}, where two metal nano-spheres support plasmonic dark modes coupled to the metal nanorod. We consider the nanorod has nonlinearity as a second harmonic generation. The dynamics of the structure as follows. The incident laser of frequency $\omega$ excites the first harmonic~(FH) mode ($\hat{b}_1$) of the nanorod and generates second harmonic (SH) one~($\hat{b}_2$). The resonances of the FH and SH modes are $ \omega_{b1} $ and $ \omega_{b2} $ respectively. Each of the FH and SH modes interact with the first~($\hat{d}_1$) and second~($\hat{d}_2$) dark modes, which have resonances at $ \omega_{d1} $ and $ \omega_{d2} $ respectively.

The components of the Hamiltonian for the coupled system in the light of the above definitions can be expressed as follows. The free oscillations of the bright and dark plasmon modes~$\hat{H}_0$ together with the excitation of the FH mode of the nanorod by the pump source~($\hat{H}_{P} $) can be given as
\begin{eqnarray}
\hat{H}_0 &=& \hbar \sum_{j=1}^2 \{ \omega_{bj} \hat{b}_j^\dagger \hat{b}_j+\omega_{dj} \hat{d}_{j}^\dagger \hat{d}_j \}, \\
\hat{H}_{P} &=& i\hbar (\varepsilon_p  \hat{b}_1^\dagger e^{-iwt} -\textit{h.c}),
\label{Ham0}
\end{eqnarray} 
where $\hat{b}_j^\dagger$ ($\hat{b}_j$) and $\hat{d}_j^\dagger$~($\hat{d}_j$) define the creation~(annihilation) of the $j$th mode of the bright and dark plasmons respectively with $j=1,2$. $\varepsilon_p $ is proportional to the laser field amplitude. Hamiltonian for the second harmonic generation process $\hat{H}_{sh} $ and the interaction between the bright and dark plasmon modes  $ \hat{H}_{int}$ are given by
\begin{eqnarray}
\hat{H}_{sh} &=& \hbar \chi^{(2)} (\hat{b}_2^\dagger \hat{b}_1 \hat{b}_1+ \hat{b}_1^\dagger \hat{b}_1^\dagger \hat{b}_2 ),\\  
\hat{H}_{int}&=&\hbar \sum_{j=1}^2 g_j \{\hat{b}_j^\dagger \hat{d}_j+ \hat{d}_j^\dagger \hat{b}_j  \}.
\end{eqnarray}
Here the parameters  $ g_1 $ and $ g_2 $ , in units of frequency, are the coupling strengths of the linear and second harmonic responses of the nanorod to the dark modes of the nanospheres respectively. $ \chi^{(2)} $ is the overlap integral, which is proportional to the second-order susceptibility of the nanorod. Thus, the total Hamiltonian can be written as: $ \hat{H}=\hat{H}_0+\hat{H}_P+\hat{H}_{sh}+\hat{H}_{int}$. The dynamics of the system can be derived by using Heisenberg equation of motion, e.g., $ i\hbar \dot{\hat{a}}_j=[\hat{a}_j, \hat{H}] $. To investigate only the field amplitudes, we replace the second-quantized operators with their expectations, e.g., $ \alpha_j=\langle \hat{b}_j \rangle $ and $ \beta_j=\langle \hat{d}_j \rangle $ with $ j=1,2 $. The resulting equations of motion by including the decay rates can be obtained as
\begin{subequations}
\begin{align}
\dot{{\alpha}}_1&=-(i\omega_{b1}+\gamma_{b1}) {\alpha}_1-i 2\chi^{(2)} {\alpha}_1^\ast {\alpha}_2-i g_1 {\beta}_1+\varepsilon_p e^{-i\omega t} \label{EOMa},\\
\dot{{\alpha}}_2&=-(i\omega_{b2}+\gamma_{b2}) {\alpha}_2-i \chi^{(2)} {\alpha}_1^2-i g_2 {\beta}_2 \label{EOMb},\\
\dot{{\beta}}_1&=-(i\omega_{d1}+\gamma_{d1}) {\beta}_1-i g_1 {\alpha}_1 \label{EOMc},\\
\dot{{\beta}}_2&=-(i\omega_{d2}+\gamma_{d2}) {\beta}_2-i g_2 {\alpha}_2 \label{EOMd},
\end{align}
\end{subequations}
where $ \gamma_{bj} $ and $ \gamma_{dj} $  stand for the decay rates of the bright and dark modes respectively. To obtain the Fano resonance in such a system, these parameters play an essential role. In definition, Fano resonance can be observed when a continuum state interacts with the discrete one, where the dark-bright coupling fits this picture as dark modes have much longer lifetimes comparing the bright ones, i.e., $ \gamma_{dj}\ll \gamma_{bj} $.

\begin{figure*}
\centering
\includegraphics[width=12cm]{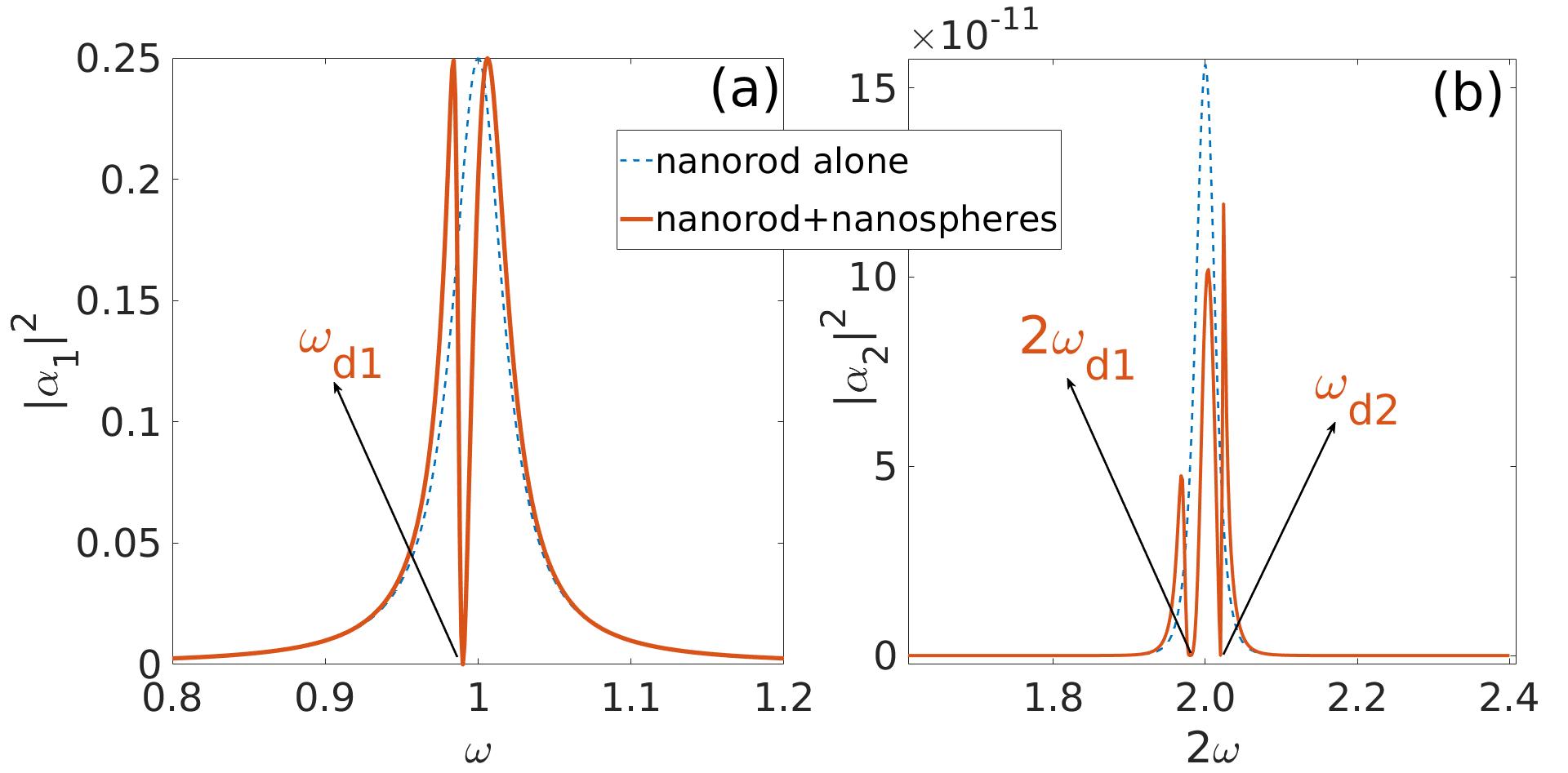}
\caption{The field intensities of the FH (a) and SH (b) when the nanorod alone (blue-dashed) and with the nanospheres~(red-straight) obtained from  the numerical solutions of the Eqs.~(\ref{EOMa}) - (\ref{EOMd}). A transparency window for the FH can be seen at $ \omega=\omega_{d1} $ that the nonlinear contribution from nonlinear Fano resonance at $ \omega=\omega_{d2}/2 $ does not change FH intensity profile. On the other hand, there are two dips for the SH intensity: one due to nonlinear Fano resonance when $ 2\omega=\omega_{d2}$ and the other due to $\tilde{\alpha}_1^2$ term in the nominator of the Eq.~(\ref{steady_2}). More details in the text.}
\label{fig2}
\end{figure*}

\begin{figure*}[h]
\centering
\includegraphics[width=12cm]{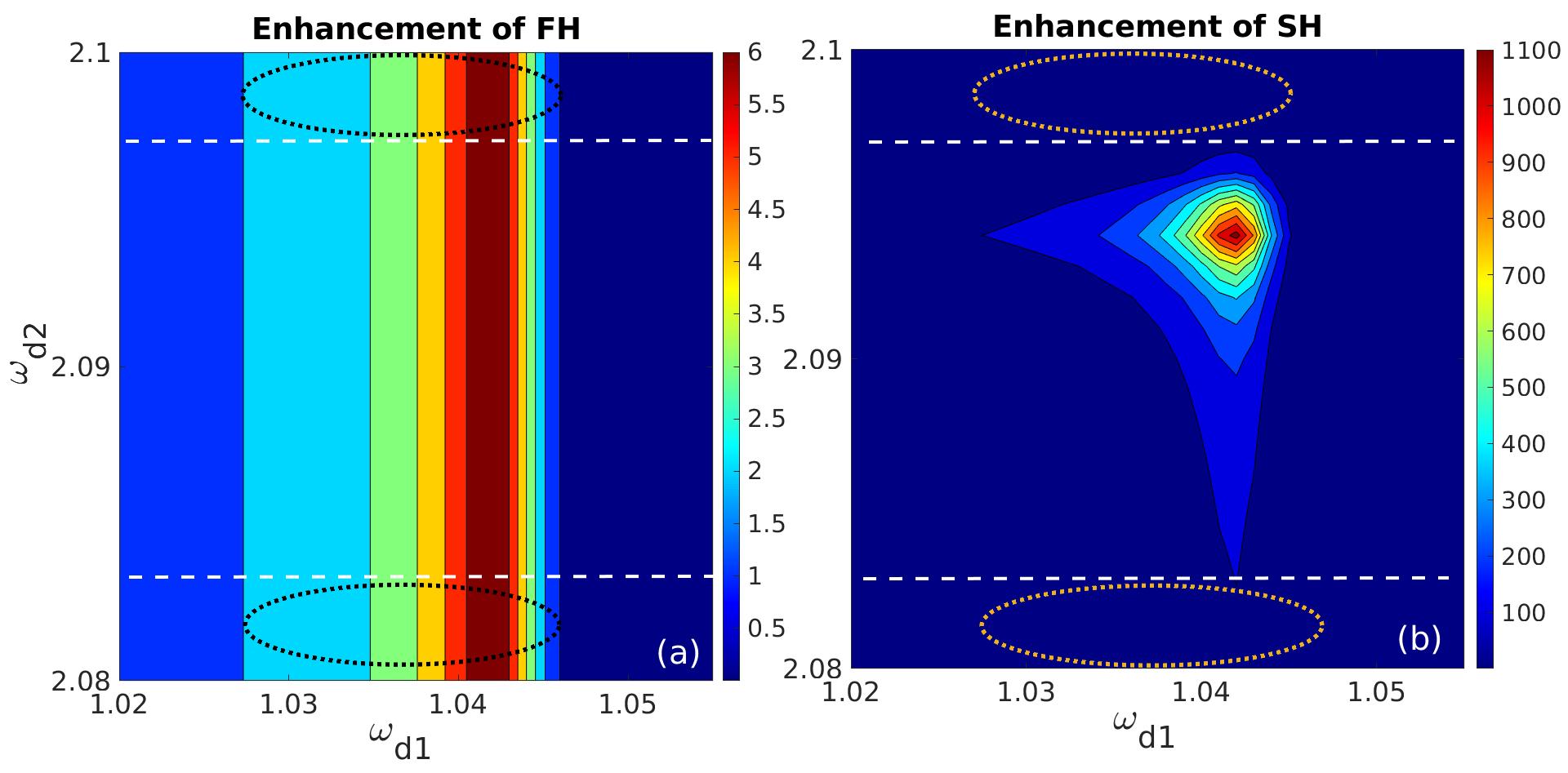}
\caption{The enhancement of the field intensities of the FH (a) and SH (b) obtained from  the numerical solutions of the Eqs.~(\ref{EOMa}) - (\ref{EOMd}). Here we take $ \omega=1.05~\omega_{b1} $ and the rest of the parameters are as given in the text. In the elliptic regions, SH field intensity is suppressed even the FH gets enhanced. In other words, nonlinear Fano resonance cancels linear Fano resonance contribution.  }
\label{fig3}
\end{figure*}

As the system is driven by a source proportinal to the term $ e^{-i\omega t} $, one can expect that the solutions at the steady-state oscillate as:
\begin{eqnarray}
\alpha_k=\tilde{\alpha}_k e^{-ik\omega t}, \qquad \beta_k=\tilde{\beta}_k e^{-ik\omega t}\qquad k:1,2,
\end{eqnarray}
where amplitudes with tilde are slowly varying terms that determine the steady-state field amplitudes. By inserting these equations into the equations of motion,  Eqs.~(\ref{EOMa}) - (\ref{EOMd}), we obtain the equations for FH~($\tilde{\alpha}_1$) and SH~($\tilde{\alpha}_2$) steady-state amplitudes as
\begin{eqnarray}
\tilde{\alpha}_1&=& \frac{\varepsilon_p}{i(\omega_{b1}-\omega)+\gamma_{b1}+\frac{g^2_1}{i(\omega_{d1}-\omega)+\gamma_{d1}}}+f(\chi^{(2)}),\label{steady_1}\\
\tilde{\alpha}_2&=& \frac{-i \chi^{(2)} \tilde{\alpha}_1^2}{i(\omega_{b2}-2\omega)+\gamma_{b2}+\frac{g^2_2}{i(\omega_{d2}-2\omega)+\gamma_{d2}}}, \label{steady_2}
\end{eqnarray}
where $f(\chi^{(2)})$ is the function proportional the $\chi^{(2)}$ term, which does not play essential role in the suppression and the enhancement of the FH field~(see Figure \ref{fig2}). Although, we obtain the Figures~\ref{fig2} and~\ref{fig3} by solving the Eqs.~(\ref{EOMa}) - (\ref{EOMd}) numerically, we study the steady-state solutions in Eqs.~(\ref{steady_1}) - (\ref{steady_2}) to get better insight for the Fano enhancement of the linear and the nonlinear fields.

The suppression occurs due to the emergence of a transparency window induced by Fano resonance, which does not allow an excitation at the converted frequency. It is shown in Figure.~\ref{fig2} that the transparency window can be obtained both in the linear and the second harmonic field intensities. Different from the FH intensity, there are two dips in the response of SH intensity as shown. This can be explained by a closer look at the denominators of the Eqs.~(\ref{steady_1}) - (\ref{steady_2}). That is when the pump frequency becomes resonant with one of the dark modes, i.e., $ \omega=\omega_{d1} $  and/or $ 2\omega=\omega_{d2} $, the second terms in these denominators become ruling as long as the interaction strength is sufficient. Unlike FH intensity, in the SH intensity there is a $ \tilde{\alpha}_1^2 $ dependence in the nominator that when $ \omega=\omega_{d1} $ is met, the dip also appears in the SH intensity due to this dependence. Since the $ \chi^{(2)}$ term, in general, is small in most of the materials comparing the other frequencies the $ f(\chi^{(2)}) $ term does not play an essential role in the FH intensity when $ 2\omega=\omega_{d2} $ is met. This can be seen in Figure.~\ref{fig2}(a). Here we scale all the frequencies with $\omega_{b1}$ and use the values $\omega_{b2}=2 \omega_{b1}$, $\gamma_{b1}=\gamma_{b2}=0.05\omega_{b1}$, $\gamma_{d1}=\gamma_{d2}=10^{-4}\omega_{b1}$ for the numerical calculations in the obtained results. As we focus on the weak coupling regime~($g_i<\gamma_{bi},\gamma_{di}$), we take the interaction strengths as $g_i=0.25\gamma_{bi}$

Similarly, the interference effect can be arranged in such a way that the nonlinear process can be carried out near the resonance. That is by playing the resonance values of the dark-modes, the denominators of the Eqs.~(\ref{steady_1}) - (\ref{steady_2}) can be minimized in which the field amplitudes can attain large values. In Figure.~\ref{fig3}, the enhancement of both FH and SH intensities can be observed. Similar to the suppression process nonlinear Fano resonance does not play a role in the FH field intensity. On the other hand, linear Fano resonance multiplies the nonlinear one that we obtain 3-orders of magnitude enhancement in the SH field intensity even the FH field intensity gets enhanced only 6 times. By the tidy arrangement of the parameters, one can obtain a better cancellation scheme and hence a larger enhancement of the nonlinear field intensity. It can be seen that, in the elliptic regions of the Figure.~\ref{fig3}, SH field intensity is suppressed even the FH gets enhanced, where the nonlinear Fano resonance cancels linear Fano resonance contribution. This result can be useful when the nonlinearity is undesired in such a system.

\section{Conclusion} \label{Conclusion}
In summary, we demonstrate that controlling second-harmonic field intensity is possible by using both linear and nonlinear Fano resonances. Our results are twofold. On the one hand, one can obtain a larger enhancement of the nonlinear field, which is at least one order of magnitude larger comparing with the enhancement obtained via only the nonlinear Fano resonance. On the other hand, linear field enhancement can be obtained by suppressing the nonlinear fields. This can be useful when the nonlinearity is undesired. Our results contribute to the understanding of the amplification of optical nonlinearities via Fano resonances and can find potential applications from all-optical switch nonlinear devices to ultrafast spectroscopies.

\end{document}